\documentclass[twocolumn,floats,showpacs,amsmath,amssymb,prb]{revtex4}

\usepackage{graphicx}

\begin{document}

\title{Isotope effects in ice Ih: A path-integral simulation}
\author{Carlos P. Herrero}
\author{Rafael Ram\'{\i}rez}
\affiliation{Instituto de Ciencia de Materiales,
         Consejo Superior de Investigaciones Cient\'{\i}ficas (CSIC),
         Campus de Cantoblanco, 28049 Madrid, Spain }
\date{\today}

\begin{abstract}
   Ice Ih has been studied by path-integral molecular dynamics 
simulations, using the effective q-TIP4P/F potential model for flexible
water.
This has allowed us to analyze finite-temperature quantum effects in this
solid phase from 25 to 300 K at ambient pressure.
Among these effects we find a negative thermal expansion of ice at low
temperatures, which does not appear in classical molecular dynamics
simulations.
The compressibility derived from volume fluctuations gives 
results in line with experimental data.
We have analyzed isotope effects in ice Ih by considering normal, heavy, 
and tritiated water. In particular, we studied the effect of changing
the isotopic mass of hydrogen on the kinetic energy and atomic
delocalization in the crystal, as well as on structural properties such
as interatomic distances and molar volume. For D$_2$O ice Ih at 100 K
we obtained a decrease in molar volume and intramolecular O--H 
distance of 0.6\% and 0.4\%, respectively, as compared to H$_2$O ice.
\end{abstract}



\maketitle

\section{Introduction}

Condensed phases of water have been studied over the years by many
experimental and theoretical techniques. Several of their properties
have been well explained, but some others lack a complete
understanding, in part due to the peculiar structure of liquid and solid
water, in which hydrogen bonds between adjacent molecules give rise to
properties somewhat different than those of most known liquids and
solids (the so-called water ``anomalies'').\cite{ei69,pe99,fr00,ro96}

The structure of ``normal'' ice Ih has been known for many years.
A feature displayed by this and some other ice phases is that although
water molecules are disposed on a regular crystal lattice, their
spatial orientation is disordered, being only controlled by the
so-called ice rules.\cite{pa35,be35}
Thus, ice Ih is not an ordinary crystal, in the sense that oxygen
atoms lay on regular lattice positions, but the hydrogen positions are
randomly chosen among the six possible orientations of each molecule,
in such a way that the orientations of neighboring molecules are
compatible.\cite{ku05}

Computer simulation of water in condensed phases attracted much
interest after the development of Monte Carlo and molecular
dynamics methods for studying liquids and solids at an atomic level. 
In the first works,\cite{ba69,ra71} rigid nonpolarizable models for the 
water molecule were employed, and subsequently a considerable amount of 
attention was focused on the development and refinement
of empirical potentials to describe both liquid and solid phases of
water. Nowadays, a large variety of this kind of potentials are present
in the literature.\cite{ko04,jo05,ab05,pa06,mc09}
Many of them assume a rigid geometry for the water molecule, and some 
others include molecular flexibility either with harmonic
or anharmonic OH stretches. Also, in some cases the polarizability
of the water molecule has been explicitly introduced into the model
potentials.\cite{ma01}
Moreover, in recent years, some simulations of water using 
\textit{ab initio} density functional theory (DFT) have appeared in the 
literature.\cite{fe06,mo08}
Nevertheless, the hydrogen bonds present in condensed phases of water
seem to be difficult to describe with presently available energy
functionals, which causes that some properties are poorly reproduced 
by DFT simulations.
This is the case of the melting temperature of ice Ih, which can be
overestimated by more than 100 K.\cite{yo09}

A limitation of {\em ab-initio} electronic-structure calculations in 
condensed matter is that they usually treat atomic 
nuclei as classical particles, and typical quantum effects like 
zero-point motion are not directly accessible.
These effects can be included by using harmonic or quasiharmonic 
approximations, but are difficult to take into account when large 
anharmonicities are present, as can happen for light atoms like 
hydrogen.
To consider the quantum character of atomic nuclei, the path-integral 
molecular dynamics (or Monte Carlo) approach has proved to be very useful.
A remarkable advantage of this procedure is that all nuclear degrees of
freedom can be quantized in an efficient manner, thus including
both quantum and thermal fluctuations in many-body systems
at finite temperatures.  In this way, Monte Carlo or molecular dynamics
sampling applied to evaluate finite-temperature path integrals allows
one to carry out quantitative and nonperturbative studies of 
highly-anharmonic effects in condensed matter.\cite{gi88,ce95}
Earlier studies of ice using path-integral simulations have been 
carried out by using mainly effective potentials, and were
focused on structural and dynamic properties of the solid 
phase.\cite{ga96b,he05,he06b,pa08,mc09}

A typical quantum effect is the isotopic dependence of several
properties of a crystal, which would not vary with the atomic masses in
a classical approach.  Thus,
the actual lattice parameters of two chemically identical crystals 
with different isotopic composition are not equal, as a consequence
of the dependence of the atomic vibrational amplitudes
on the atomic mass.\cite{bu88,ho91,no97}
Lighter isotopes have larger vibrational amplitudes
(as expected in a harmonic approximation) and larger
lattice parameters (an anharmonic effect).
This effect is most noticeable at low temperatures,
as the atoms in the solid feel the anharmonicity
of the interatomic potential due to zero-point motion.
At higher temperatures, the vibrational amplitudes are larger
(causing a larger volume), but the isotope
effect on the crystal volume becomes less prominent, because
those amplitudes are less mass-dependent. In the high-temperature
(classical) limit this isotope effect disappears.
Something similar happens for the interatomic distances in the solid.
Path integral simulations have been used earlier to study isotopic
effects in solids. In particular, this technique has turned out to be
sensitive enough to quantify the dependence of crystal volume on the
isotopic mass of the constituent atoms.\cite{no97,he99,he09c}  

All these quantum effects become more relevant as the atomic mass
decreases, and will be particularly important in the
case of hydrogen. Then, we pose the question of how the mass of the
lightest atom can influence the structural properties of a solid water
phase, in particular ice Ih. This refers to the crystal volume, but
also to the interatomic distances in the solid. Moreover, changing the
hydrogen mass should modify the actual binding energy of the crystal,
since the kinetic energy of lighter atoms will be higher.  

With this purpose, we study in this paper ice Ih by path-integral 
molecular dynamics (PIMD) simulations. This technique allows us to analyze 
in a quantitative way several effects associated to the quantum nature 
of atomic nuclei, an in particular the influence of isotopic mass on 
the crystal volume and the interatomic distances. 
We consider normal (H$_2$O), heavy (D$_2$O), and tritiated (T$_2$O) water.
Interatomic interactions are described by the flexible q-TIP4P/F model, 
which has been recently developed and was employed to carry out PIMD 
simulations of liquid water.\cite{ha09}  In an earlier paper\cite{ra10}
 we have used this interatomic potential to study the melting of 
ice, by using nonequilibrium techniques which allowed us to obtain the
coexistence temperature of solid and liquid water phases at ambient
pressure. Here we employ equilibrium PIMD to compare results for ice, 
obtained for the different hydrogen isotopes.

 The paper is organized as follows. In Sec.\,II, we describe the
computational method and the models employed in our calculations. 
Our results are presented in Sec.\,III, dealing with atomic
delocalization, kinetic energy, crystal volume, 
interatomic distances, and bulk modulus of ice
Ih.  Sec.\,IV includes a summary of the main results.

\section{Computational Method}

\subsection{Path-integral molecular dynamics}

Our calculations are based on the path-integral formulation of 
statistical mechanics. In this formulation, the partition function 
of a quantum system is evaluated by a discretization of the 
density matrix along cyclic paths, consisting  of a finite number 
$L$ (Trotter number) of steps.\cite{fe72,kl90} 
In the implementation of this technique to numerical simulations, 
such a discretization gives rise to the appearance of 
$L$ replicas (or beads) for each quantum particle. These replicas can 
be treated in the calculations as classical particles, since the 
partition function of the original quantum system is isomorph to 
that of a classical one, obtained by replacing each quantum particle by a 
ring polymer made of $L$ particles.\cite{gi88,ce95}
This path-integral approach allows one to study finite-temperature
properties of quantum many-body problems in a nonperturbative
scheme, even in the presence of large anharmonicities.\cite{ce95}
The configuration space can be adequately sampled by
molecular dynamics or Monte Carlo techniques. Here, we have used the
PIMD method, mainly because in this case the codes are more easily
parallelizable, a decisive factor for efficient use of modern
computer architectures.    

Simulations of ice Ih have been carried out here in the 
isothermal-isobaric $NPT$ ensemble ($N$, number of particles; 
$P$, pressure; $T$, temperature), 
which allows us to calculate the equilibrium volume of the
solid at given pressure and temperature.
We have employed effective algorithms for performing PIMD simulations 
in this statistical ensemble, as those described in the 
literature.\cite{ma96,tu02}
Sampling of the configuration space has been carried out at ambient
pressure ($P$ = 1 bar) and temperatures between 25 and 300 K.  
To study isotope effects on the volume and other properties of ice,
we have considered H$_2$O, D$_2$O, and T$_2$O. This is easily done in
PIMD simulations, since the atomic mass is an input parameter in the
calculations.
For comparison, some simulations of ice Ih were also
performed in the classical limit, which is obtained in our context by
setting the Trotter number $L = 1$.
Also for the sake of comparison with the results for actual ice Ih,
we carried out some PIMD simulations in which the hydrogen mass
was assumed to be very large, so that for our purposes H behaves as 
a classical particle. In fact, we took a mass $m_{\text H}$ = 10,000 u,
which gives results indistinguishable from those obtained for even
larger masses.

Our simulations were carried out on ice Ih supercells with periodic
boundary conditions.
To check the influence of finite-size effects on the results,
we considered two kinds of orthorhombic supercells.
The smaller one included 96 water molecules and had parameters 
$(3a, 2 \sqrt{3} a, 2c)$,
where $a$ and $c$ are the standard hexagonal lattice parameters of
ice Ih.
The larger supercell included 288 molecules and had parameters
$(4a, 3 \sqrt{3} a, 3c)$.
Results obtained for both types of supercells coincided within error
bars caused by the statistical uncertainty associated to the simulation
procedure.
Prior to the PIMD simulations, proton-disordered ice structures 
were generated by a Monte Carlo procedure, in such a way that
each oxygen atom has two chemically bonded and two H-bonded hydrogen
atoms, and with a cell dipole moment close to zero.\cite{bu98}

For the interatomic interactions we have employed the point charge,
flexible q-TIP4P/F model,\cite{ha09} which has been previously used
to study liquid water, the coexistence between the
liquid and ice Ih phases,\cite{ra10} as well as water
clusters.\cite{go10}
The Lennard-Jones-type interaction appearing in the intermolecular
potential was truncated at $r_c$ = 8.5 \AA. To compensate for this
truncation, standard long-range corrections were computed assuming
that the pair correlation function $g(r)$ is unity for distance 
$r > r_c$, leading to well-known corrections for the pressure and
internal energy.\cite{jo93}
The electrostatic energy and associated long-range forces were
calculated by the Ewald method, using a parallel code to improve
the speed of the procedure.\cite{ra10}

 For a given temperature, a typical simulation run consisted of 
$4 \times 10^4$ PIMD steps for system equilibration, followed by 
$6 \times 10^5$ steps for the calculation of ensemble average properties.
To keep roughly a constant precision in the PIMD results
at different temperatures, the Trotter number was scaled with the 
inverse temperature ($L \propto 1/T$), so that $L T$ = 6000~K, 
which translates into $L$ = 20 and 240 for  $T$ = 300 K and 25 K,
respectively.

The simulations were performed by using a staging transformation
for the bead coordinates.
The constant-temperature ensemble was generated by coupling chains 
of four Nos\'e-Hoover thermostats to each staging variable.
To generate the $NPT$ ensemble, an additional chain of four barostats 
was coupled to the volume.\cite{tu98}
To integrate the equations of motion, we employed
a reversible reference-system propagator algorithm (RESPA), which allows
one to define different time steps for the integration of fast and slow
degrees of freedom.\cite{ma96} 
The time step $\Delta t$ associated to the calculation of q-TIP4P/F forces
was taken in the range between 0.1 and 0.3 fs, which was found to be 
appropriate for the interactions, atomic masses, and temperatures 
considered here, and provided adequate convergence for the studied
variables.
For the evolution of the fast dynamical variables, associated to the
thermostats and harmonic bead interactions, we used a smaller
time step $\delta t = \Delta t/4$.
Note that for ice at 25 K, a simulation run consisting of 
$6 \times 10^6$ PIMD steps requires calculation of energy and forces 
with the q-TIP4P/F potential for $1.4 \times 10^9$ atomic (classical) 
configurations, which was carried out by parallel computing.

\subsection{Spatial delocalization}

We define here some spatial properties of the nuclear quantum paths 
that are helpful in the analysis of the PIMD simulation results.
The center-of-gravity (centroid) 
of the quantum paths of a given particle is calculated as
\begin{equation}
   \overline{\bf r} = \frac{1}{L} \sum_{i=1}^L {\bf r}_i  \, ,
\label{centr}
\end{equation}
${\bf r}_i$ being the position of bead $i$ in the associated ring
polymer.

The mean-square displacement of a quantum particle along a PIMD
simulation run is then given by:
\begin{equation}
\Delta^2_r =  \frac{1}{L} \left< \sum_{i=1}^L 
           ({\bf r}_i - \left< \overline{\bf r} \right>)^2
           \right>    \, ,
\label{delta2}
\end{equation}
where $\langle ... \rangle$ indicates a thermal average at temperature $T$.
After a straightforward transformation, one can write $\Delta^2_r$ as
\begin{equation}
 \Delta^2_r = Q^2_r + C^2_r  \, ,
\label{delta2b}
\end{equation}
with
\begin{equation}
 Q^2_r = \frac{1}{L} \left< \sum_{i=1}^L 
             ({\bf r}_i - \overline{\bf r})^2 \right>    \, ,
\end{equation}
and
\begin{equation}
  C^2_r = \left< \overline{\bf r}^2 \right> -
             \left< \overline{\bf r} \right>^2     \, .
\label{cr2}
\end{equation}
The first term in Eq.~(\ref{delta2b}), $Q^2_r$, is the mean-square 
``radius-of-gyration'' of the ring polymers associated to the quantum 
particle under consideration, and gives the average spatial
extension of the paths.\cite{gi88}
The second term on the r.h.s. of Eq.~(\ref{delta2b}), $C^2_r$, 
is the mean-square displacement of the path centroid.
In the high-temperature (classical) limit, each path collapses onto 
a single point and the radius-of-gyration vanishes, i.e., $Q^2_r \to 0$.
In cases where the anharmonicity is not very large, the spatial
distribution of the centroid $\overline{\bf r}$ is similar to that 
of a classical particle moving in the same potential.

\section{Results}

\subsection{Hydrogen delocalization}

\begin{figure}
\vspace{-1.0cm}
\hspace{-0.5cm}
\includegraphics[width= 9cm]{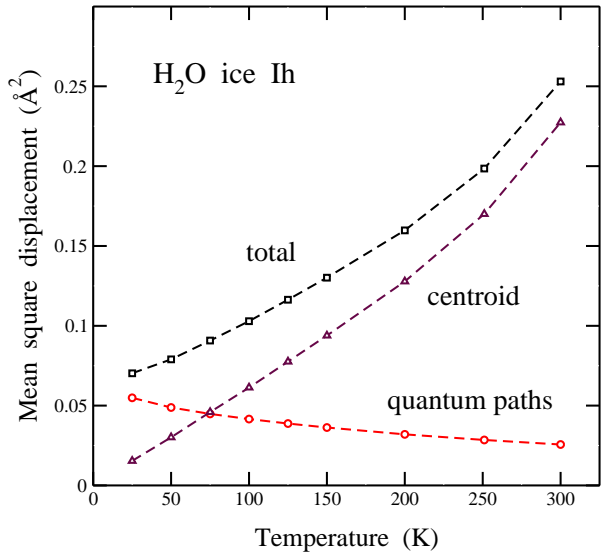}
\vspace{-1.0cm}
\caption{
Spatial delocalization of hydrogen nuclei (protons) in H$_2$O ice Ih,
as derived from PIMD simulations at several temperatures.
Symbols indicate the mean-square displacement
for the centroid $C_r^2$ (triangles), and
the radius-of-gyration of the quantum paths $Q_r^2$ (circles).
The total delocalization $\Delta_r^2$ is shown by open squares.
Lines are guides to the eye.
}
\label{f1}
\end{figure}

We first consider the spatial delocalization of hydrogen 
in normal ice Ih, which is expected to include a nonnegligible quantum
contribution.  We have calculated separately
both terms giving the atomic delocalization in Eq.~(\ref{delta2b}),
for each hydrogen atom in the water molecules.
In Fig.~1 we display the values of $Q_r^2$ (spreading of the quantum
paths, circles) and $C_r^2$ (centroid delocalization, triangles),
as derived from our PIMD simulations for the H$_2$O molecule at several
temperatures.
The total spatial delocalization $\Delta_r^2$ is shown as squares.
In this plot, one observes that at low temperatures the quantum
contribution (spreading of the paths), $Q_r^2$, dominates the spatial
delocalization of hydrogen, since the centroid displacement, $C_r^2$, 
converges to zero as $T \to 0$ K.
The opposite happens at temperatures higher than 70 K, and in the
high-temperature limit (unreachable here for stability reasons), the
quantum contribution $Q_r^2$ would eventually disappear, as corresponds
to the classical limit.
At $T$ = 250 K, close to the melting temperature predicted by this
potential model,\cite{ra10} we found
$Q_r^2 = 2.85 \times 10^{-2}$ \AA$^2$ vs $C_r^2 = 0.170$ \AA$^2$,
i.e., the former contributes a 14\% to the total spacial delocalization
of hydrogen, $\Delta_r^2$.
 We note that at low temperatures the quantum paths associated to
hydrogen have an average extension of about 0.15 \AA,
much smaller than the H--H distance in a water molecule, thus justifying 
the neglect of quantum exchange between protons in the PIMD simulations.

The mean-square displacement $\Delta_r^2$ of hydrogen atoms derived from
our PIMD simulations with the q-TIP4P/F potential are similar to those
derived by Tanaka and Mohanty\cite{ta02} from classical molecular
dynamics simulations with the TIP4P potential, in the temperature range
considered by these authors (from 150 to 250 K). At lower temperatures,
classical simulations should give values smaller than the PIMD
simulations, due to its neglect of atomic zero-point motion. 
The atomic mean-square displacements are usually supposed to be related
with the melting of a solid. Thus, the Lindemann criterion gives
a threshold for the maximum amplitude of atomic vibrations that can
be sustained by a crystal. At the melting point of ice Ih we find from
our PIMD simulations that this amplitude is about 6\% of the H-bond 
distance, as discussed elsewhere.\cite{ra10}

\begin{figure}
\vspace{-1.0cm}
\hspace{-0.2cm}
\includegraphics[width= 9cm]{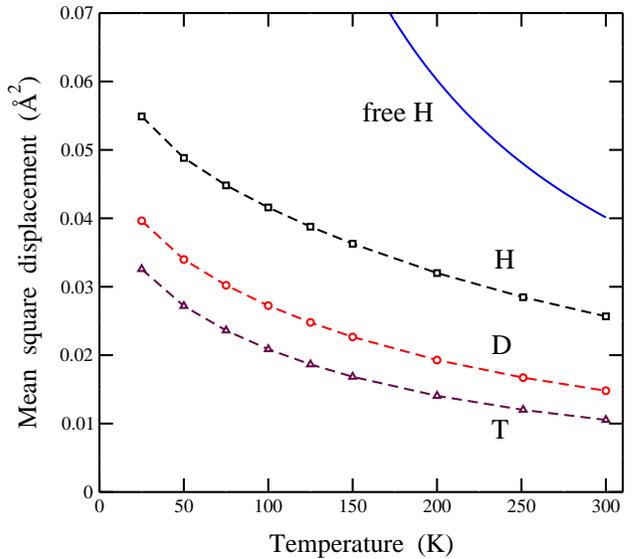}
\vspace{-1.0cm}
\caption{
Mean square radius-of-gyration of the quantum paths, $Q_r^2$,
for hydrogen isotopes in ice Ih, as derived from PIMD
simulations at several temperatures for H$_2$O (squares),
D$_2$O (circles), and T$_2$O (triangles).
Error bars are in the order of the symbol size.
The solid line represents $Q_r^2$ for a free hydrogen atom.
}
\label{f2}
\end{figure}

Quantum effects are more relevant as the mass 
of the particles under consideration becomes smaller. Thus, one expects
that the quantum delocalization of the hydrogen isotopes in ice will
decrease for rising isotopic mass. This is shown in Fig.~2, where we
have plotted the mean-square displacement of the quantum paths,
$Q_r^2$, for hydrogen (squares), deuterium (circles), and tritium
(triangles) at several temperatures.
In the zero-temperature limit and in a harmonic approximation, 
$Q_r^2$ is known to scale as $m_{\text H}^{-1/2}$.
At 25 K, we find a ratio $Q_r^2$(H)/$Q_r^2$(D) = 1.39, slightly smaller 
than the low-temperature harmonic expectancy of $\sqrt{2}$.
In the high-temperature limit $Q_r^2$ goes to zero,
but the ratio $Q_r^2$(H)/$Q_r^2$(D) converges to the inverse mass
ratio,\cite{gi88,ra93} in this case $m_{\rm D}/m_{\rm H}$ = 2.
At 300 K we found $Q_r^2$(H)/$Q_r^2$(D) = 1.73, between the high and
low-temperature limits.  
For T$_2$O we found at 300 K, $Q_r^2 = 1.05 \times 10^{-2}$ \AA$^2$,
so that $Q_r^2$(H)/$Q_r^2$(T) = 2.44, also between a ratio of
$\sqrt{3}$ expected at low temperature in a harmonic approach,
and the high-temperature limit given by $m_{\rm T}/m_{\rm H}$ = 3.

For comparison with the results of our PIMD simulations for
different hydrogen isotopes in ice Ih, we also present in Fig.~2 the
mean-square radius of gyration corresponding to a free hydrogen atom 
(without any external potential, solid line). This can be calculated
analytically, as shown elsewhere,\cite{gi88,ra93} and gives values of 
$Q_r^2$ clearly higher than for hydrogen in ice, in the whole
temperature range considered here, confirming the importance of the 
hydrogen environment in its quantum delocalization. This is even
clearly appreciable at the melting temperature.

For the centroid delocalization, $C_r^2$, of deuterium and tritium in
ice Ih we found values almost indistinguishable from those obtained for 
hydrogen (shown in Fig.~1)
in the whole temperature region under consideration, which in turn were
practically the same as the mean-square displacement of hydrogen found 
in classical molecular dynamics simulations.

\subsection{Kinetic energy}

A typical quantum effect related to the atomic motion in solids is
that the kinetic energy $E_k$ at low temperature converges to a finite value
associated to zero-point motion, contrary to the classical result where
$E_k$ vanishes at 0 K.
Path integral simulations allow one to obtain the kinetic energy
of the quantum particles under consideration, which is basically
related to the spread of the quantum paths. In fact, for a particle at
given temperature and isotopic mass, the larger the mean-square 
radius-of-gyration of the paths, $Q_r^2$, the smaller the kinetic energy.
Here we have calculated $E_k$ by using the so-called virial
estimator, which has an associated statistical uncertainty lower
than the potential energy of the system.\cite{he82,tu98}

\begin{figure}
\vspace{-1.0cm}
\hspace{-0.2cm}
\includegraphics[width= 9cm]{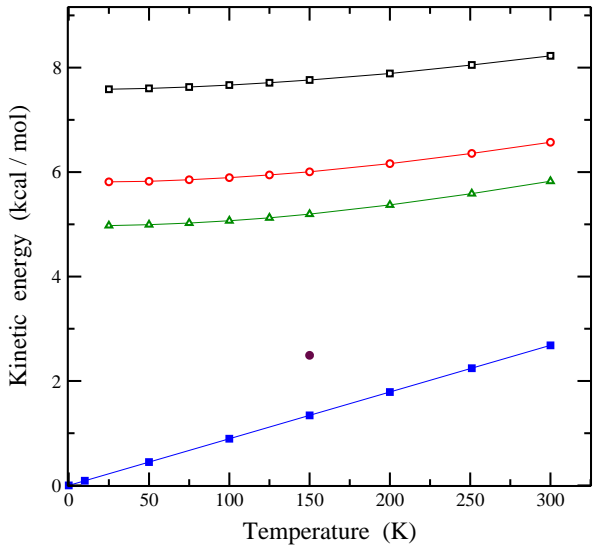}
\vspace{-1.0cm}
\caption{
Temperature dependence of the kinetic energy of ice Ih, as derived
from PIMD simulations.
Open symbols indicate results derived from simulations with full
quantum motion of the atoms in the cell: squares for H$_2$O,
circles for D$_2$O, and triangles for T$_2$O.
For comparison we also present results of classical molecular dynamics
simulations (filled squares).
The filled circle corresponds to a PIMD simulation of ice with the
hydrogen mass going to infinity.
Error bars are less than the symbol size.
Lines are guides to the eye.
}
\label{f3}
\end{figure}

In Fig.~3 we display the kinetic energy as a function of temperature
for ice Ih. Open symbols indicate results derived from PIMD
simulations: H$_2$O (squares), D$_2$O (circles), and T$_2$O
(triangles). In each case, $E_k$ increases as temperature
rises, and at low temperature it converges to the value corresponding 
to zero-point motion. At a given temperature, $E_k$ decreases
for increasing isotopic mass. Thus, at $T$ = 25 K we obtained 
$E_k$~= 7.59, 5.81, and 4.98 kcal/mol for H$_2$O, D$_2$O, and T$_2$O, 
respectively. The change in kinetic energy caused by isotopic
substitution decreases as temperature is raised. However, we
stress that at the melting temperature, the kinetic energy of
deuterated ice is still more than 1.5 kcal/mol lower than that of
normal ice, reflecting the importance of quantum effects in ice Ih at
this temperature.

For comparison with the results of our PIMD simulations, 
we also present in Fig.~3 the kinetic energy
corresponding to a classical model with $3 N$ degrees
of freedom ($N$ atoms): $E_k^{\text cl} = 3 N k_B T / 2$, solid squares.
For rising temperature,
the classical kinetic energy approaches the results of the PIMD
simulations, in particular those corresponding to the heaviest 
molecule, T$_2$O. However, at 300 K the classical value is still
lower than $E_k$(T$_2$O) by 3.14 kcal/mol, which represents about
50\% of the quantum value.
To assess the importance of quantum effects associated to the oxygen
motion, we also show in Fig.~3 the kinetic energy obtained in PIMD
simulations at 150 K for the limit $m_{\text H} \to \infty$. 
In this case we
obtain $E_k$ = 2.49 kcal/mol, i.e. 1.15 kcal/mol above the classical
limit at this temperature. This represents about 30\% of the increase
in kinetic energy from the classical limit to T$_2$O quantum ice.

\subsection{Crystal volume}

\begin{figure}
\vspace{-1.0cm}
\hspace{-0.2cm}
\includegraphics[width= 9cm]{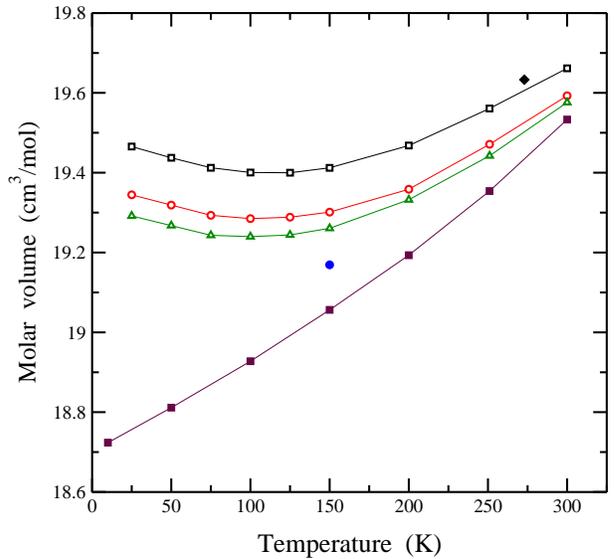}
\vspace{-1.0cm}
\caption{
Molar volume of ice Ih as a function of temperature, as derived
from PIMD simulations for H$_2$O (squares), D$_2$O (circles),
and T$_2$O (triangles).
For comparison, results of classical molecular dynamics simulations
(filled squares) are also shown.
A filled circle corresponds to the ice volume obtained from a PIMD
simulation with the hydrogen mass going to infinity.
Error bars are in the order of the symbol size.
Lines are guides to the eye.
A solid diamond represents the experimental determination from
Refs.~\onlinecite{gi47,da68} at 273 K.
}
\label{f4}
\end{figure}

As commented in the Introduction, the crystal volume is expected to
change with the isotopic mass of the constituent atoms. We have
calculated the equilibrium volume of ice Ih in our
isothermal-isobaric PIMD simulations for different hydrogen masses.
In Fig.~4 we present the temperature dependence of the 
molar volume for the different isotopes.  Results of our PIMD
simulations at $P$ = 1 bar are shown as open symbols: H$_2$O
(squares), D$_2$O (circles), and T$_2$O (triangles).
One first notices that the change of volume with temperature is not
monotonic, but at low temperatures it decreases as temperature is raised,
to reach a minimum at $T \approx 100$ K. At higher temperature, the
molar volume increases, and the solid recovers the usual thermal
expansion.  Also, the volume of the solid decreases as the isotope
mass of hydrogen is raised, as a consequence of the smaller 
spatial (quantum) delocalization associated to a larger nuclear mass.
We observe for all three isotopes the same behavior of the volume,
with a minimum around 100 K.
The volume changes derived from the PIMD simulations are isotropic,
i.e., the ratio between parameters of the simulation cell is constant
when changing the temperature, within the statistical error bar,
which agrees with experimental data.\cite{ro94}
For thermodynamic reasons, one expects that in the low-temperature
limit ($T \to 0$ K) the thermal expansion of the solid goes to zero,
i.e., $d V / d T \to 0$. We cannot, however, check this point at
present since PIMD simulations at temperatures lower than 25 K would 
require enormous computational resources. 

The molar volume (or density) of ice has been determined experimentally
in very different ways, e.g., calorimetric, acoustical, mechanical,
x-ray, optical, or nuclear methods (see \onlinecite{fe06b} and
references therein). Measurements of different authors typically
deviate from each other by up to about 0.3\%, which could be caused by
the density lowering effect of aging on ice crystals or by their very
slow relaxation to equilibrium.\cite{fe06b} 
The molar volumes 19.635 cm$^3$/mol of Ginnings and Corruccini\cite{gi47} 
and 19.633 cm$^3$/mol of Dantl and Gregora\cite{da68} are considered 
in Ref.~\onlinecite{fe06b} the most accurate determinations at normal 
pressure and 273 K. They are plotted in Fig.~4 as a solid diamond.
At lower temperatures, there appears some dispersion in the
experimental points presented by different authors.
A minimum in the molar volume was found in several
works\cite{da62,ro94,fe06b} in the range between 60 and 90 K, to be
compared with the minimum at around 100 K derived from our PIMD
simulations. The minimum molar volume was found to be around 
19.3 cm$^3$/mol, somewhat smaller than our calculated value at 100 K
(19.40 cm$^3$/mol). This, along with the rather good agreement between
calculated and experimental values at the melting temperature,
indicates that the employed interatomic potential underestimates the 
thermal expansion of the solid.
In this context, the most remarkable point is that the q-TIP4P/F
potential is able to capture the negative thermal expansion 
of ice at low temperatures.
This phenomenon is associated to the tetrahedral coordination
of the water molecules in ice, and has been observed in other
solids with similar structures. In particular, it is well known for 
crystals with diamond and zinc blende structure, such as Si, GaAs, and
CuCl.\cite{ev99}

Another important point to be stressed here is the appreciable
isotopic effect on the molar volume. In fact, at  $T$ = 100 K,
we found $v$ = 19.400, 19.285, and 19.240 cm$^3$/mol, for
H$_2$O, D$_2$O, and T$_2$O, respectively, which translates into
a volume reduction of 0.6\% and 0.8\% in passing from
normal ice to deuterated and tritiated ice, respectively. 
From the crystal structure obtained by R\"ottger {\em et al.}\cite{ro94}
one derives for normal ice Ih at 100 K a molar volume of
19.301 cm$^3$/mol, somewhat smaller than that found in the PIMD
simulations with the q-TIP4P/F potential. 
These authors also measured the lattice parameters of deuterated ice,
and found an inverse isotopic effect, which is not reproduced in
our simulations.

For comparison, we also present in Fig.~4 the molar volume of ice Ih 
derived from classical molecular dynamics simulations with the same
interatomic potential (filled squares). At a given 
temperature, the volume obtained from the classical simulations is
smaller than that found in quantum PIMD simulations, but the
difference decreases as $T$ rises.
Note that, contrary to the results of the quantum simulations,
the molar volume obtained in the classical ones increases
monotonically and does not show any anomaly in the whole temperature 
region up to 300 K.
At $T$ = 100 K, we obtain in the classical approximation a
molar volume $v$ = 18.928 cm$^3$/mol. This means that at this
temperature, quantum effects cause for normal ice Ih a remarkable
volume expansion of about 2.5\%.
In the zero-temperature limit one can estimate for H$_2$O ice a volume
increase of about 0.8 cm$^3$/mol, which means a rise around 4.3\%
due to nuclear quantum effects. 

We note that both the thermal expansion (positive or negative) and
the isotopic effect on the crystal volume are anharmonic effects that
would be absent in a harmonic approximation.
On one side, the thermal expansion appears in both classical and
quantum simulations, since in both cases the atomic vibrations feel
the anharmonicity of the interatomic potential at any finite
temperature ($T >$ 0 K). In the quantum case, the
anharmonicity is also felt in the low-temperature limit $T \to 0$ K,
due to zero-point motion, and thus in this limit the crystal volume is
larger for quantum than for classical calculations (which may be
called ``zero-point crystal expansion''). 
On the other side, the isotopic effect on the crystal volume is a
typical quantum effect that disappears in the classical limit, and is
associated to the different vibrational amplitudes of different
isotopes (which coincide in the classical limit).

The negative thermal expansion of ice at low $T$ is due to low-energy 
transverse vibrational modes with negative Gr\"uneisen parameter.
For a mode in the $n$'th phonon branch with wave vector ${\bf q}$, this
parameter $\gamma_n({\bf q})$ is defined from the logarithmic 
derivative of its frequency with respect to the crystal volume:\cite{as76}
\begin{equation}
\gamma_n({\bf q}) = - \frac {\partial \ln \omega_n({\bf q})} 
 {\partial \ln V} \, . 
\end{equation}
At relatively low temperature, low-energy modes are more populated than 
modes with higher energy (with positive $\gamma_n({\bf q})$), and then
the overall contribution to the volume change with increasing
temperature will be negative.\cite{ta98,be99b}
In this line, Tanaka calculated the temperature dependence of the crystal
volume in a quasiharmonic approximation and obtained a negative thermal
expansion at low temperatures.\cite{ta98,ta01} 
This author calculated the relative
weight of different vibrational modes to the overall thermal expansion
at different temperatures, and found that a family of low energy modes
($\omega <$ 50 cm$^{-1}$), corresponding to bending motion of three 
water molecules is largely responsible for the negative thermal
expansion at low temperatures.
In connection with this, an instability in a transverse acoustical mode
in ice Ih was found at low temperature from incoherent inelastic
neutron scattering.\cite{be99b} 

We remember that in the classical molecular dynamics simulations of ice
Ih we did not obtain any negative thermal expansion, in contrast
with PIMD simulations. This is in line with the explanation given
above related to the different weight of phonons with different energies
at low temperatures. In a classical model all phonons have the same
weight at any temperature (equipartition principle), and thus the
negative Gr\"uneisen parameter of low-energy transverse modes is largely
compensated for at any temperature by the positive $\gamma$ of most 
vibrational modes in the solid. 

It is worthwhile commenting on the effect of pressure upon the molar
volume of ice. It is well known that ice Ih is no longer mechanically
stable at pressures in the order of 1 GPa, where it has been observed
to transform into an amorphous phase.\cite{mi84,sc95} 
This has been in fact obtained in
our PIMD simulations at $T =$ 251 K and $P =$ 1 GPa, where the ice
crystal collapsed into an amorphous solid, with a volume decrease of
about 18\%. At $P =$ 0.8 GPa and the same temperature we found that 
the ice crystal was still stable along our simulation runs, and 
obtained a reduction in the crystal volume of 8.1\% with respect to 
ambient pressure. Interestingly, at 0.8 GPa the atomic delocalization 
of hydrogen was found to be $\Delta_r^2$ = 0.309 \AA$^2$, to be
compared with a value of 0.199 \AA$^2$ obtained at $P =$ 1 atm
(see above), which means a large increase in the mean-square
displacement of about 50\%.
Such an anomalous increase in the atomic delocalization seems to be
related with the crystal instability associated to the amorphization 
of ice Ih under pressure. This question requires further investigation
by using path-integral simulations at different pressures.

\subsection{Interatomic distances}

\begin{figure}
\vspace{-1.0cm}
\hspace{-0.2cm}
\includegraphics[width= 9cm]{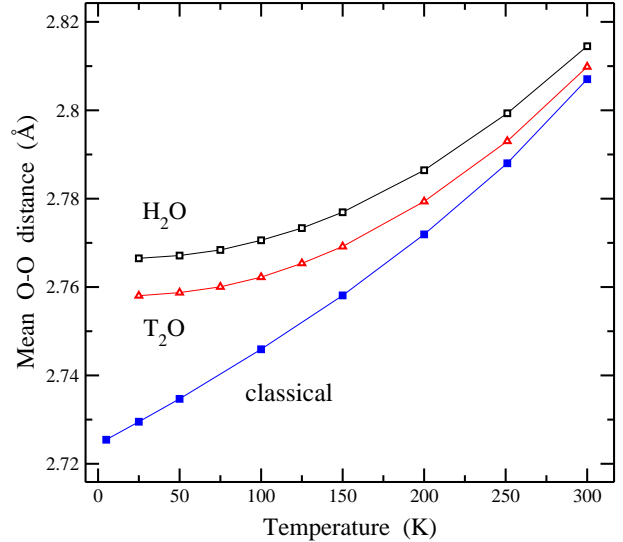}
\vspace{-1.0cm}
\caption{
Mean distance between oxygen atoms in nearest-neighbor water
molecules, as a function of temperature. Open symbols correspond
to results of PIMD simulations for H$_2$O (squares) and
T$_2$O (triangles). Filled squares represent results of classical
molecular dynamics simulations.
Lines are guides to the eye.
}
\label{f5}
\end{figure}

In this subsection we present results for interatomic distances
in ice Ih between atoms in the same and adjacent molecules.
We first show in Fig.~5 the mean distance between oxygen atoms in 
neighboring molecules. Open symbols represent results of PIMD
simulations: squares for H$_2$O and triangles for T$_2$O.
Data points for the case of D$_2$O lie between those of H$_2$O
and T$_2$O, and are not shown for clarity.
For comparison we also present in this figure the average O--O
distance derived from classical molecular dynamics simulations
(filled squares).
Comparing results of PIMD and classical simulations, we find that 
in the low-temperature limit the O--O distance for normal ice 
increases by about 0.04 \AA\ with respect to the classical value,
i.e., around 1.5\%.
The O--O distance derived from quantum simulations increases as 
temperature is raised, and the slope rises from a vanishing value 
at $T \to 0$ K to nearly the classical value at 300 K.
For tritiated ice we found an interatomic distance between that 
for H$_2$O ice and the classical limit.

 For a totally homogeneous and isotropic crystal expansion 
one expects a volume change $\Delta V / V = 3 \Delta l / l$, $l$ being 
any distance in the solid. This relation is valid for small volume
changes, i.e., for $\Delta V / V \ll 1$, as occurs in the temperature
range considered here, and in particular, one would expect it
to be fulfilled for the intermolecular distance in the solid.
Thus, from the increase in the average O--O distance at low temperature 
for ice Ih one expects a volume expansion of about 4.5\% due to
zero-point motion. This value is close to that estimated from the
volumes obtained in classical and quantum simulations at low
temperature, which give an expansion of 4.3\% (see above, Sec.~III.C).
Note, however, that a fixed relation between interatomic distances and 
crystal volume
can not be strictly followed in the case of ice Ih when the temperature 
is raised, specially in the region where the crystal shrinks.
In fact, we observe that the average O--O distance does 
not present a minimum at $T \approx$ 100 K, as the molar volume of ice.
This clearly reflects the fact that shrinking of the ice crystal is not 
due to a reduction in the distance between nearest molecules, but to a
bending motion of contiguous molecules, corresponding to
low-frequency vibrational modes, as indicated above in Sec.~III.C.
From the results of our quantum and classical simulations
at 100 K we find a volume increase of 2.5\% due to quantum nuclear
effects vs a relative rise of 0.9\% in the average O--O distance,
which would correspond to a volume change of 2.7\% for an isotropic and
homogeneous expansion. 

\begin{figure}
\vspace{-1.0cm}
\hspace{-0.2cm}
\includegraphics[width= 9cm]{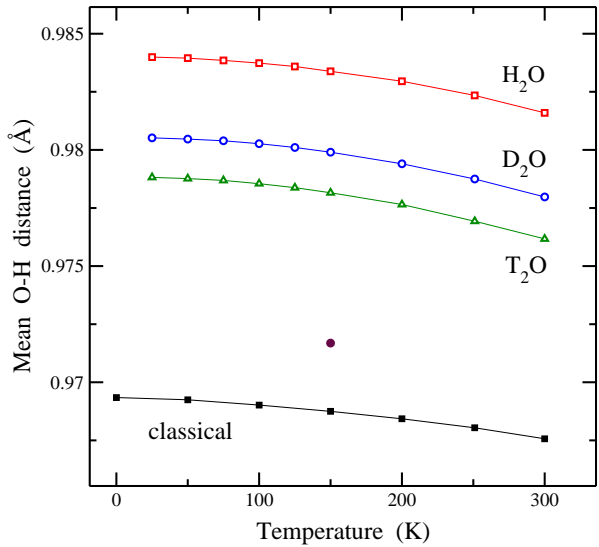}
\vspace{-1.0cm}
\caption{
Temperature dependence of the mean intramolecular O--H distance,
as derived from simulations of ice Ih. Open symbols represent
results of PIMD simulations for H$_2$O (squares), D$_2$O (circles),
and T$_2$O (triangles).
For comparison, the average O--H distance obtained from classical
molecular dynamics simulations (filled squares) are also presented.
 The filled circle
corresponds to the limit $m_{\text H} \to \infty$ in PIMD simulations.
Error bars are in the order of the symbol size.
Lines are guides to the eye.
}
\label{f6}
\end{figure}

We now turn to analyze the interatomic distances inside water molecules
in the ice crystal, as derived from our simulations at different
temperatures.
In Fig.~6 we have plotted the mean intramolecular O--H distance 
as a function of temperature. As in previous figures, open
symbols indicate results of PIMD simulations for H$_2$O (squares),
D$_2$O (circles), and T$_2$O (triangles). 
At a given temperature, the average O--H distance is larger
for smaller hydrogen isotopic mass. 
In particular, the difference between O--H and O--D distances amounts to
about $4 \times 10^{-3}$ \AA\ in the whole temperature range under
consideration, which agrees with data derived from Compton scattering
experiments for water at room temperature.\cite{ny07}
For comparison, we present in Fig.~6 results for the classical limit
(filled squares), which lie clearly lower than those of the quantum
simulations. To assess the relative importance of quantum nuclear
effects of hydrogen and oxygen, we give also the O--H distance obtained
in a PIMD simulation in the limit of infinite hydrogen mass (filled
circle at 150 K). This point is still clearly above the classical
limit, with a quantum contribution (coming from quantum delocalization
of oxygen nuclei)
amounting to 31\% and 20\% of the increase in O--H distance for 
full quantum T$_2$O and H$_2$O ice, respectively.

It is interesting that the intramolecular O--H distance decreases
for increasing temperature, in both classical and quantum approaches.
This fact has been observed
experimentally and reported in the literature.\cite{ny06}
It is due to the peculiar structure of ice with hydrogen bonds
connecting neighboring water molecules. 
As temperature is raised, molecular motion is enhanced, so that 
hydrogen bonds become softer, and the average intermolecular O--H 
distance rises (in agreement with the increase in O--O distance
presented above). This causes a strengthening of the intramolecular 
O--H bonds, with a concomitant decrease in the interatomic distance in
water molecules.
The same behavior is found from the classical molecular dynamics 
simulations (filled squares in Fig.~6),
although in this case the mean intramolecular O--H distance is
clearly smaller than in the full quantum ice.

\begin{figure}
\vspace{-1.0cm}
\hspace{-0.5cm}
\includegraphics[width= 9cm]{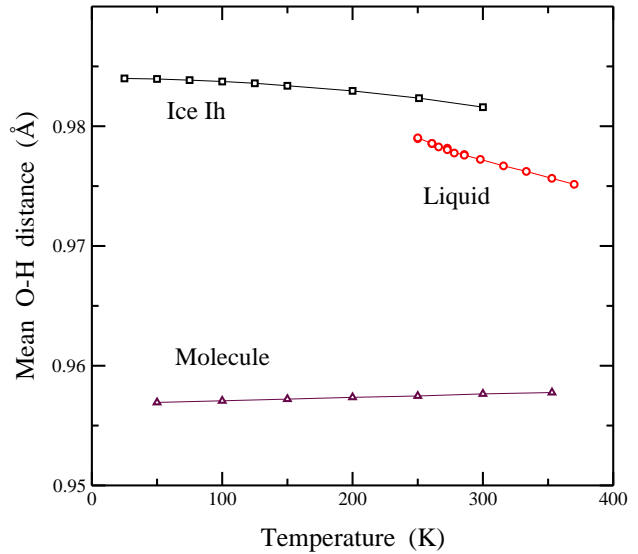}
\vspace{-1.0cm}
\caption{
Temperature dependence of the average intramolecular O--H distance,
as derived from PIMD simulations of a single H$_2$O molecule
(triangles), ice Ih (squares), and liquid water (circles).
Error bars are in the order of the symbol size.
Lines are guides to the eye.
}
\label{f7}
\end{figure}

To obtain deeper insight into the contraction of the intramolecular
O--H bond for rising temperature, we compare in Fig.~7 the average O--H
distance derived from our PIMD simulations for ice Ih (open squares) 
with those obtained for liquid water (circles) and for an isolated 
H$_2$O molecule (triangles).   
One observes first that the O--H distance in the water molecules
increases appreciably in passing from the gas phase to condensed phases
(solid and liquid). This is a consequence of hydrogen bonds with
the nearest molecules, which weaken the intramolecular bonds.  
The enlargement of these intramolecular bonds is larger for the solid
than for the liquid, since in the former the intermolecular hydrogen
bonds are stronger due to the lower mobility of the molecules and the
larger rigidity of the H-bond network.
In the case of ice Ih, the bond enlargement with respect to an isolated
H$_2$O molecule amounts to about 0.03 \AA,
which means a 3\% of the bond distance.
Second, one observes that the O--H distance for an isolated
molecule increases slightly as the temperature is raised, contrary to
the condensed phases, where this distance decreases for increasing  
temperature.
The behavior observed for the isolated molecule is the usual thermal 
expansion due to anharmonicities in the interatomic potential.
For the condensed phases, this natural expansion is largely
compensated for by intermolecular interactions through hydrogen
bonds, as indicated above. As a result, the intramolecular O--H
distance in these phases decreases as the temperature rises.

\subsection{Bulk modulus} 

Among the many surprising properties of liquid water and ice, one finds
that their compressibility is smaller than what one could naively expect
from the large cavities present in their structure, which could
presumably collapse under pressure without water molecules approaching 
close enough to repel each other. For ice Ih, in particular, the hydrogen 
bonds holding the crystal structure are rather stable, as manifested
by the relatively high pressure necessary to break down the H-bond 
network and amorphize the solid.\cite{sc95}

The isothermal compressibility $\kappa$ of ice, or its inverse the bulk 
modulus [$B = 1/\kappa = - V ( {\partial P} / {\partial V} )_T$] can be 
calculated directly from our PIMD simulations in the isothermal-isobaric 
ensemble. In fact, in the $NPT$ ensemble the isothermal bulk modulus
can be obtained from the mean-square fluctuations of the volume,
$\sigma_V^2 = \langle V^2 \rangle - \langle V \rangle^2$,
by using the expression\cite{la80,he08}
\begin{equation}
       B = \frac{k_B T \langle V \rangle}{\sigma_V^2}   \; ,
\label{bulkm}
\end{equation}
where $k_B$ is Boltzmann's constant.
This expression has been employed earlier to calculate the bulk modulus
of different types of solids from path-integral simulations.\cite{he01,he08}
Here, we have used Eq.~(\ref{bulkm}) to obtain $B$ for ice Ih from
both classical and PIMD simulations.

In the classical simulations we find that the bulk modulus decreases 
roughly linearly as temperature is raised, and in the zero-temperature
limit it extrapolates to a value $B_0 = 17.6 \pm 0.3$ GPa.
In the quantum simulations of H$_2$O ice Ih, the solid is found to be 
``softer'' than in the classical simulations, in the sense that in 
the former the volume fluctuations are larger, and consequently the 
bulk modulus becomes smaller. Note that in Eq.~(\ref{bulkm}) one also
has the average volume $\langle V \rangle$ in the numerator, which is
larger for the quantum than for the classical solid (see Sec.~III.C),
but this volume increase due to quantum nuclear effects is dominated by
the also present larger volume fluctuations $\sigma_V^2$. 
In the low-temperature limit we find an extrapolated value of
$B = 13.8 \pm 0.4$ GPa, about 3.8 GPa smaller than the classical value,
which means that the quantum effect reduces the bulk modulus by more than
25\%.

Note that the values of $B$ derived from our simulations in the 
isothermal-isobaric ensemble show relative error bars larger than those
corresponding to other calculated variables (e.g., kinetic energy,
interatomic distances, or molar volume).  This is a consequence of
the statistical uncertainty of the volume fluctuations $\sigma_V$, that
are employed to calculate the bulk modulus $B$. The error bars are
similar for both classical and PIMD simulations.

\begin{figure}
\vspace{-1.0cm}
\hspace{-0.2cm}
\includegraphics[width= 9cm]{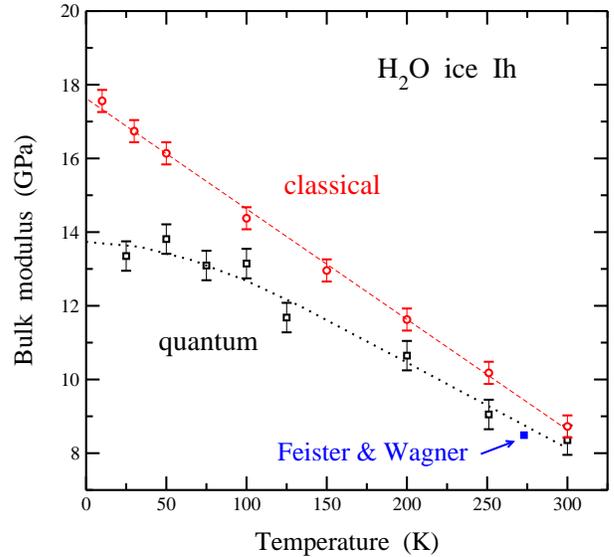}
\vspace{-1.0cm}
\caption{
Bulk modulus of ice Ih as obtained from quantum PIMD (open squares)
and classical molecular dynamics (open circles) simulations for
H$_2$O.
Error bars show the statistical uncertainty in the values of $B$
found from the simulations.
The filled square corresponds to the value obtained at 273 K from the
equation of state derived by Feistel and Wagner\cite{fe06b} from
experimental data.  Lines are guides to the eye.
}
\label{f8}
\end{figure}

There appears in the literature a dispersion of data for the isothermal
compressibility (or bulk modulus) of ice Ih at the melting temperature
and normal pressure. 
The bulk modulus of ice found from our PIMD simulations is close to 
that derived by Feistel and Wagner,\cite{fe06b} who obtained an
equation of state for ice Ih from a set of experimental data. 
This value at 273 K is shown in Fig.~8 as a solid square.
It is hard to estimate the precision of this value of the bulk modulus 
due to lack of error bars for the published data, and the
uncertainty may be large, as suggested by the dispersion in
the experimental data obtained by different authors.
The uncertainty is even larger at lower temperatures, where we could
not find direct experimental data. From the equation of state
by Feistel and Wagner\cite{fe06b} one derives at low temperature and
normal pressure an isothermal bulk modulus of 10.6 GPa, somewhat lower
than the low-temperature extrapolation of our simulation results. 

The bulk moduli for D$_2$O and T$_2$O ice derived from our PIMD
simulations show a similar trend to
that of normal ice Ih, and have not been presented in Fig.~8 for
clarity. They seem to lie between the classical limit and
the quantum value for H$_2$O, but due to the data dispersion, the
difference between $B$ values for the different hydrogen isotopes can 
be hardly quantified.     Moreover,
we do not observe in the bulk modulus of ice any anomaly similar to
that of the crystal volume at $T \approx 100$ K, but we cannot exclude it
from our present results because of their statistical uncertainty.

\section{Summary}

We have presented results of PIMD simulations of ice Ih with different
hydrogen isotopes in a wide range of temperatures.
This technique allows us to explore readily isotope effects, since the 
atomic masses appear as input parameters in the calculations.
This kind of simulations enable one to calculate separately kinetic and 
potential energies at finite temperatures, taking into account the
quantization of nuclear motion. 
This includes consideration of zero-point motion of the atoms in the
solid, which can be hard to analyze by analytical calculations in the
presence of light atoms and large anharmonicities.

This kind of quantum simulations are necessary to reproduce some
observed facts of ice, that are not captured by classical simulations.
In this sense, PIMD simulations with the q-TIP4P/F potential are able
to reproduce the negative thermal expansion of ice Ih at low
temperatures. Also, these simulations reproduce the apparently
anomalous decrease of the intramolecular O--H distance for increasing
temperature. A good check of the employed interatomic potential is the
calculation of a macroscopic observable such as the bulk modulus, which
has been accurately reproduced in our calculations.
Note that the use of a flexible potential model for water has allowed us 
to look in a realistic way at changes in the intramolecular O--H distance 
and quantum delocalization of hydrogen at different temperatures. 

PIMD simulations give reliable predictions for several isotope
effects in condensed matter. For ice Ih, in particular, we have seen
how the isotopic mass of hydrogen affects the kinetic energy and atomic
delocalization in the crystal, as well as the interatomic
distances and molar volume. Thus, we found for D$_2$O ice Ih at 100 K a
decrease in the crystal volume and intramolecular O--H distance of 0.6\%
and 0.4\%, respectively, as compared to H$_2$O ice. 
Our simulations, however, do not reproduce the inverse isotopic effect
in the crystal volume, observed in Ref.~43.

Simulations similar to those presented here can be carried out for ice
under an external pressure, and in particular to study the amorphization
process of the ice crystal. Also, the question of quantum tunneling and
diffusion of hydrogen in the different ice structures needs further
investigation.
These are challenging problems that could be addressed in the near
future by using path-integral simulations.

\begin{acknowledgments}
This work was supported by Ministerio de Ciencia e Innovaci\'on (Spain)
through Grants FIS2006-12117-C04-03 and FIS2009-12721-C04-04,
and by Comunidad Aut\'onoma de Madrid through Program
MODELICO-CM/S2009ESP-1691.
\end{acknowledgments}

\bibliographystyle{apsrev}

\end{document}